\documentstyle[12pt]{article}
\begin{document}
\thispagestyle{empty}
\parskip=4mm
\parindent=0cm
\newcommand{\bra}{\langle\,}
\newcommand{\ket}{\,\rangle}
\newcommand{\lsim}{\,\hbox{\rlap{\raisebox{-0.8ex}{$\sim$}}\raisebox{0.3ex}{$<$}}\,}
\newcommand{\R}{{\mathchoice{\hbox{$\sf\textstyle I\hspace{-.15em}R$}}
{\hbox{$\sf\textstyle I\hspace{-.15em}R$}}
{\hbox{$\sf\scriptstyle I\hspace{-.10em}R$}}
{\hbox{$\sf\scriptscriptstyle I\hspace{-.11em}R$}}}}
\renewcommand{\theequation}{\arabic{section}.\arabic{equation}}
\renewcommand{\d}{\displaystyle}
\vfill{UNIL-TP-3/95}
\hfill{hep-ph/9509217}

\vspace{20mm}
\begin{center}
{\Large\bf Lowest tensor-meson resonances contributions\\[2mm]
to the chiral perturbation theory\\[4mm]
low energy coupling constants}
\vspace{8mm}

D. Toublan

{\small Institut de physique th\'eorique, Universit\'e de Lausanne\\
CH-1015 Lausanne, Switzerland\\
DTOUBLAN@ULYS.UNIL.CH

\vspace{19mm}

{\bf Abstract}

\parbox{14cm}{The contributions of the lightest tensor-meson resonances
to the low-energy coupling constants of second order chiral
perturbation theory with two flavours are evaluated and compared with the
available phenomenological information as well as with similar results for
other resonances.}}
\end{center}

\vfill{PACS numbers : 13.75.Lb, 14.40.Cs, 12.39.Fe.}

\setcounter{page}{0}
\newpage
\section{Introduction}
The physics of pions at low energy can be described by an effective Lagrangian
given by the chiral perturbation theory (ChPT). This Lagrangian depends on
some low-energy coupling constants which are fixed in principle by the
underlying QCD, although their exact determination is still inaccessible.

In the framework of 2nd-order ChPT~\cite{Gas}, the contributions to the
coupling constants of all meson resonances of spin $\leq 1$ have been
established and their importance has been shown~\cite{Eck}. In fact, it is the
vector-meson which dominates, but the others cannot be ignored. The
tensor-meson resonance has a higher mass than those analyzed in~\cite{Eck}
but, because of its multiplicity, it is hard to tell whether its contribution
is trifling or not.

The $SU(2)$ ChPT to second order in the momenta and the external fields is
presented briefly in Section~2. The kinematics of a spin-2 field is depicted in
Section~3 using a construction made by Fierz and Pauli~\cite{Fie}. In
Section~4, the coupling of the tensor-meson resonances with the pions and their
contribution to the low-energy coupling constants is shown, following the
general trend used in~\cite{Eck}, and our way of describing the tensor-meson is
compared with the one developped in~\cite{Bell}. Section~5 is devoted to a
comparison of Pauli-Fierz scheme with other ones and finally in Section~6 a
general overview of the contributions of the different meson-resonances and a
comparison with the phenomenological values of the coupling constants are
given.

\setcounter{equation}{0}
\section{Second-order ChPT with two flavours}

To make the presentation simpler, we shall only give those results we consider
necessary here, and the reader is invited to consult~\cite{Gas} or~\cite{Eck}
for a more detailed discussion.

ChPT is an effective theory describing QCD at low energies. The two-flavours
massless-quarks QCD Lagrangian is symmetric under $SU(2)_R\times SU(2)_L$, the
chiral group. It is assumed that a spontaneous chiral symmetry breakdown
occurs,
\[SU(2)_R\times SU(2)_L \to SU(2)_V,\]
whose Goldstone bosons are identified as the pions.

The QCD Lagrangian can be approximated at a given order in the momentum using
an effective Lagrangian expressed in terms of a field $U\in SU(2)$ which
transforms linearly under $SU(2)_R\times SU(2)_L$,
\[ U\to g_RUg_L^+,\]
and contains the fields of the three pseudoscalar Goldstone bosons,
\begin{equation}
U=u^2={\rm e}^{{\rm i}\phi^a\tau_a/F},\qquad \phi^a\tau_a=
\left(\begin{array}{cc}
\pi^0&\sqrt{2}\pi^+\\ \sqrt{2}\pi^- &-\pi^0\end{array}\right),
\end{equation}
where $F$ is the pion decay constant in the chiral limit:
$F_\pi=F(1+O(m_{\mbox{\tiny quark}}))$.

Coupling $U$ with external fields and expanding the effective Lagrangian in
powers of the external momenta and of quark masses, gives
\begin{equation}
{\cal L}_{\rm eff}={\cal L}_2+{\cal L}_4+\dots
\end{equation}
The local nature of the chiral symmetry requires the introduction of a
derivative ${\rm D}_\mu U$ which is covariant with respect to the external
vectorial and axial gauge fields.

It is found that ${\cal L}_2$ depends on two parameters and ${\cal L}_4$ on
seven:
\begin{equation}
{\cal L}_4=\sum_{i=1}^7 l_i \; P_i,
\end{equation}
where the terms $P_i$ contain external fields, pion fields and their
derivatives of order $p^4$. In particular, it will be shown that only the
first two parameters receive contributions from the tensor-meson resonance; the
corresponding $P_i$ are:
\begin{eqnarray}
P_1&=&{1\over 4}\bra u^\mu u_\mu\ket^2 \\
P_2&=&{1\over 4}\bra u_\mu u_\nu\ket\,\bra u^\mu u^\nu\ket,
\end{eqnarray}
using the notation of~\cite{Bell},
\[u_\mu={\rm i}u^\dagger{\rm D}_\mu Uu^\dagger=-{\rm i}u{\rm D}_\mu U^\dagger
u=u_\mu^\dagger, \]
and $\bra \dots\ket$ stands for the trace.

The generating functional of second-order ChPT consists of the trees and
one-loop graphs generated by ${\cal L}_2$ and the trees of ${\cal L}_4$. The
divergences of the one-loop
functional are absorbed by the $l_i$, hence they will depend on a
renormalization scale $\mu$, which drops out in all observable quantities.
Denoting the renormalization parameters by $l_i^r(\mu)$, scale-independent
parameters $\bar{l_i}$ can be defined:
\begin{equation}
l_i^r(\mu)={\gamma_i\over 32 \pi^2}\left(\bar{l}_i+\ln{m^2\over \mu^2}\right),
\end{equation}
where $\gamma_i$ is a given real number~\cite{Gas} and $m$ is the pion mass.

\setcounter{equation}{0}
\section{Kinematics of a spin-2 field}

The description of a spin-2 field requires a tensor-field $T_{\mu\nu}$ that is
symmetric. However, as this object still contains 10 degrees of freedom, we
add further that its trace should vanish and that $\partial^\mu T_{\mu\nu}$
should obey a simple condition implied by the classical equations of motion,
to get the simplest interaction term arising when the coupling of the
tensor-field with the Goldstone bosons of ChPT are considered.

The attempt to obtain the equations of motion from the most general quadratic
Lagrangian by a variational principle using the holonomic constraints
$T_{\mu\nu}=T_{\nu\mu}$ and $T_\mu^\mu=0$ produces a condition on
$\partial^\mu T_{\mu\nu}$ which is not very simple. For this reason we prefer
to follow~\cite{Fie} and introduce an auxiliary scalar field $C$, independent
of $T_{\mu\nu}$ which gives a simple additional condition on
$\partial^\mu T_{\mu\nu}$ derived by variation from the Lagrange function. We
take $C$ and $T_{\mu\nu}$ to be real fields.

Denoting the mass of the tensor-field by $M$ and using Fierz and Pauli's
choice, the Lagrangian becomes
\begin{eqnarray}
{\cal L}_T&=& -{1\over 4}M^2T_{\mu\nu}T^{\mu\nu}+{1\over 4}
\partial_\rho T^{\mu\nu}\,\partial^\rho T_{\mu\nu}-{1\over 2}\partial^\rho
T_{\rho\mu}\,\partial_\nu T^{\nu\mu}\nonumber\\
&&\qquad +{3\over 16}M^2C^2-{3\over 32}\partial_\rho C\,\partial^\rho C
+{1\over 4}\partial^\mu T_{\mu\nu}\,\partial^\nu C\\ [4mm]
&&\qquad +J_{\mu \nu} T^{\mu \nu}+CJ, \nonumber
\end{eqnarray}
where $T_{\mu\nu}=T_{\nu\mu}$ and $T_\mu^\mu=0$; $J_{\mu \nu}$ and $J$ are
external fields.

Taking the second derivative of the classical equation of motion for
$T_{\mu\nu}$, with respect to $x_\mu$ and $x_\nu$, together with the classical
equation of motion for $C$, gives a linear system of equations for
$T_S\equiv \partial^{\mu\nu}T_{\mu\nu}$ and $C$:
\begin{equation}
{\cal M}\left(\begin{array}{c} T_S\\C\end{array}\right)=
\left(\begin{array}{c} j_1\\j_2\end{array}\right),
\end{equation}
where
\begin{equation}
{\cal M}=\left(\begin{array}{cc} M^2-{1\over 2}\mbox{\large $\Box$} &
{3\over 8}\mbox{\large $\Box$}\mbox{\large $\Box$}\\
-4 & 3\mbox{\large $\Box$} +6M^2\end{array}\right)
\end{equation}
and
\begin{equation}
\left\{\begin{array}{rcl}
j_1 &=& 2(\partial^{\mu\nu}J_{\mu \nu}-{1\over 4}
\mbox{\large $\Box$}J_\rho^\rho)\\
j_2 &=& -16 J.
\end{array}\right.
\end{equation}

Because $\det({\cal M})=6M^4\neq 0$, (3.2) can be inverted:
\begin{eqnarray}
\left(\begin{array}{c} T_S\\C\end{array}\right)
&=&{\cal M}^{-1}\left(\begin{array}{c} j_1\\j_2\end{array}\right). \\
%\mbox{with}\qquad{\cal M}^{-1}&=&{1\over 6M^4}
%\left(\begin{array}{cc} 3\mbox{\large $\Box$} +6M^2 &-{3\over 8}
%\mbox{\large $\Box$}\mbox{\large $\Box$}\\
%4 & M^2-{1\over 2}\mbox{\large $\Box$}\end{array}\right).
\nonumber
\end{eqnarray}
Introducing (3.5) into the classical equation of motion of $T_{\mu\nu}$,
differentiated once with respect to $x_\mu$, gives
\begin{equation}
\partial^\mu T_{\mu\nu}={1 \over 4M^2} \left( 8\partial_\mu J_{\mu \nu} - 2
\partial_\nu J_\rho^\rho + 2\partial_\nu j_1-{1 \over 4} \mbox{\large $\Box$}
\partial_\nu j_2  \right).
\end{equation}
The equation of motion becomes
\begin{eqnarray}
\left(\mbox{\large $\Box$}+M^2\right)T_{\mu\nu}&=&\Theta_{\mu\nu} \\
\hspace{-1cm} \mbox{\rm where } \Theta_{\mu\nu}&=&2(J_{\mu \nu}
-{1\over 4}g_{\mu\nu}J_\rho^\rho) \nonumber \\
&& \quad +\frac{1}{6M^4}( 4\partial_\mu \partial_\nu -g_{\mu\nu} \mbox{\large
$\Box$} ) \, (j_1-{1 \over 8} (\mbox{\large $\Box$}+M^2 ) j_2 ) \\
&& \quad +\frac{1}{M^2}(2 \partial_\mu \partial^\rho J_{\rho\nu} + 2
\partial_\nu \partial^\rho J_{\rho\mu} - \partial_{\mu\nu} J_\rho^\rho -{1
\over 2}g_{\mu\nu} j_1). \nonumber
\end{eqnarray}

Thus the effective action ca be written as:
\begin{eqnarray}
S_T&=& {1 \over 2}\int{\rm d}x\,{\rm d}y\, \left[
J^{\mu\nu}(x)\,\Pi_{\mu\nu\rho\sigma}(x-y)\Theta^{\rho\sigma}(y)+
j_1(x) K_1(x-y) j_1(y) \right. \nonumber \\
&& \qquad \left. +j_1(x) K_{12}(x-y) j_2(y)+j_2(x) K_2(x-y) j_2(y) \right] \\
[4mm]
\mbox{with }\Pi_{\mu\nu\rho\sigma}(x)&=&\int{{\rm d}^4k\over
(2\pi)^4}{{\rm e}^{-{\rm i}kx}\over k^2-M^2+{\rm
i}\epsilon}\left[{1 \over 2}(\bar{g}_{\mu\rho}(k) \bar{g}_{\mu\sigma}(k) +
\bar{g}_{\mu\sigma}(k) \bar{g}_{\mu\rho}(k)) \right. \nonumber \\
&& \hspace{8cm} \left. -{1 \over 3} \bar{g}_{\mu\nu}(k)
\bar{g}_{\rho\sigma}(k)\right], \nonumber \\
\hspace{-1cm} \bar{g}_{\mu\nu}(k)&=&g_{\mu\nu}-\frac{k_\mu k_\nu}{M^2},
\nonumber \\
\hspace{-1cm} K_1(x) &=& -\delta^4(x) {1 \over 36M^8} \left( 3 \mbox{\large
$\Box$}+6M^2\right), \nonumber \\
\hspace{-1cm} K_{12}(x) &=& -\delta^4(x) \left( {1\over 24M^4}+
{1 \over 144M^8} (-3\mbox{\large $\Box$} \mbox{\large $\Box$} + 6M^4) \right),
\nonumber \\
\hspace{-1cm}\mbox{and } K_2(x) &=& \delta^4(x) \left( {1\over 96M^4}+{1 \over
384M^8} \mbox{\large $\Box$} \mbox{\large $\Box$} \right)
(M^2-{1 \over 2}\mbox{\large $\Box$}). \nonumber
\end{eqnarray}
The coupling with the auxiliary scalar field $C$ does not generate poles.

Switching off the external fields, the Lagrangian~(3.1) describes the
kinematics of a spin-2 field using a tensor-field $T_{\mu\nu}$ which
satisfies the conditions
\begin{equation}
\left\{ \begin{array}{rcl} T_{\mu\nu} &=& T_{\nu\mu},\\
g^{\mu\nu} T_{\mu\nu} &=& 0,\\
\partial^\mu T_{\mu\nu} &=&0,
\end{array}\right.
\end{equation}
with the usual equations of motion,
\begin{equation}
\left(\mbox{\large $\Box$} +M^2\right)T_{\mu\nu}=0.
\end{equation}

\setcounter{equation}{0}
\section{Chiral coupling of the tensor-meson resonance and its contribution to
the ChPT Lagrangian}

As with the other meson resonances treated in~\cite{Eck}, the tensor-meson
effect on second-order ChPT generates a contribution to some low-energy
coupling constants.

All resonances carry non-linear realizations of the chiral group
$SU(2)_R\times SU(2)_L$ which depend on their transformation properties under
the subgroup $SU(2)_V$.

The lightest spin-2 resonances are around the same energy~\cite{Par}:
$f_2(1270)$, $a_2(1320)$, which is not coupled with two pions but with three
because of $G$-parity, $f'_2(1525)$, that mainly decays into $K \bar{K}$
($\Gamma(f'_2 \to \pi\pi) \cong 0.6$~MeV), and $\pi_2(1670)$, which has again
wrong $G$-parity to interact with two pions.

Hence the lightest spin-2 field which has a significant coupling
with the pions is the singlet $f_2(1270)$ ($J^{PC}=2^{++}$)~\cite{Par},
the realization of $SU(2)_R\times SU(2)_L$ is trivial
\[T_{\mu\nu}\;\stackrel{SU(2)_R\times SU(2)_L}{\mbox{\Large $\longrightarrow$}}
\;T_{\mu\nu}.\]

To obtain the contributions of the tensor-meson resonance to the low-energy
coupling constants, the lowest-order couplings in the chiral expansion are
needed. These are linear in the resonance fields and at most of order $p^2$.

Because of the symmetries and of the null trace property of $T_{\mu\nu}$, the
only interaction that can occur is with the external current
\begin{equation}
J_{\mu\nu}=K_T\bra u_\mu u_\nu\ket.
\end{equation}
The coupling constant $K_T$ determines the width,
\begin{equation}
\Gamma(f_2\to\pi\pi)={K_T^2M^3\over 40\pi F^4}\left(1-{4m^2\over
M^2}\right)^{5\over 2}.
\end{equation}
Using the experimental results \footnote[1]
{\begin{tabular}{rl}
&$F = 93.2$~MeV  is the pion decay constant;\\
&$m=140$~MeV is the pion mass;\\
&$M=1275$~MeV is the $f_2$ mass;\\
and & $\Gamma(f_2\to\pi\pi)=157$~MeV is the measured width~\cite{Par}
\end{tabular} }
, equation (4.2) gives
\begin{equation}
|K_T|=28.5\mbox{ MeV.}
\end{equation}
The external scalar current $J$ contains all possible coupling respecting
the symmetries of the effective chiral Lagrangian. Since the pions cannot
interact at rest, $J$ must be of order $p^2$.

At order $p^4$ the Lagrangian~(3.1) may
receive polynomial corrections that can be expressed in terms of the $P_i$
of~(2.3) in order to be consistent with QCD~\cite{Leut}.
Only the $P_i$ with the right symmetry have to be taken into account.
Moreover the terms generated by the auxiliary field $C$ in~(3.9) at this order
are precisely of this type. Hence both can be treated together.

On the one hand, using~(3.9) and~(4.1), the $\pi^0 \pi^0$ scattering amplitude
in the Mandelstam variables reads:
\begin{eqnarray}
T^{\pi^0 \pi^0}(s,t,u)&=&{8K_T^2 \over F^4} \left[ {1 \over M^2-s} \Biggl\{
-5m^4+3m^2s-{1\over 3}{m^4 s \over M^2}-{1 \over 4}s^2+{1 \over 2} (t^2+u^2) -
{1 \over 2} {m^2s^2 \over M^2} \right. \nonumber \\
&& \quad \left. -{1 \over 6} {s^3 \over M^2} +{1 \over 6}{s^3 \over M^2} +{1
\over 6}{m^2 s^3 \over M^4} +{1 \over 12}{s^4 \over M^4} \Biggr\} +{1 \over
M^2-t}  \Biggl\{ (s,t,u) \to (t,u,s) \Biggr\}
\right. \nonumber \\
&& \quad \left. + {1 \over M^2-u}  \Biggl\{ (s,t,u) \to (u,s,t) \Biggr\}
\right] + {\rm polynomial}.
\end{eqnarray}
Due to the Froissart bound, a once-subtracted fixed $t$ dispersion relation can
be written for this amplitude:
\begin{equation}
T^{\pi^0 \pi^0}(s,t,u)=\mu(t)+{1 \over \pi}\int_{4m^2}^\infty{\rm d}x\, {1
\over x^2} \left( {s^2 \over x-s} +  {u^2 \over x-u} \right) {\rm Im}T^{\pi^0
\pi^0}(s,t,u),
\end{equation}
implying that for high $s$ the contribution of the tensor-meson to the forward
$\pi^0 \pi^0$ scattering amplitude must be a constant.
Then~(4.4) is made compatible with QCD by subtracting $K_T^2$ $P_1/3$ to the
Lagrangian~(3.1).

On the other hand, the tensor-meson cannot contribute to the scalar form-factor
of the pion, which is the case for the corrected Lagrangian. This imply that
there is no other correction involving the remaining $P_i$.

The corrected Lagrangian, $\bar{{\cal L}}_T$, is now compatible with QCD. It
will be used to compute the contributions of the tensor-meson to the low-energy
coupling constants of ChPT.

With the external current~(4.1), the equation of motion reads
\begin{equation}
\left(\mbox{\large $\Box$}+M^2\right)T_{\mu\nu}=2K_T\left(\bra u_\mu u_\nu\ket
-{1\over 4}g_{\mu\nu}\bra u_\mu u^\mu\ket\right) +S_{\mu\nu},
\end{equation}
where $S_{\mu\nu}=O(p^4)$ and contains only pion fields, scalar and
pseudoscalar external fields and their derivatives (typically,
$\bra {\rm D}_\mu {\rm D}^\tau {\rm D}_\tau(U^2)^\dagger{\rm
D}^\mu(U^2)\ket$).
Since we only want the Lagrangian at order $p^4$, $T_{\mu\nu}$ only needs to
be known at order $p^2$. Equation~(4.6) implies that
\begin{equation}
T_{\mu\nu}={2K_T\over M^2}\left(\bra u_\mu u_\nu\ket-{1\over 4}g_{\mu\nu}\bra
u_\tau u^\tau\ket\right) +O(p^4)
\end{equation}
and the corrected action at order $p^4$ is
\begin{eqnarray}
\bar{S}_T&=& \int{\rm d}^4x\,\bar{{\cal L}}_T^{(4)}(x)+O(p^6),\nonumber\\[4mm]
\bar{{\cal L}}_T^{(4)}(x)&=&-{4K_T^2\over 3M^2}\, \left( {1\over 4}
\bra u_\rho u^\rho\ket^2\right)+ {4K_T^2\over M^2}\,
\left({1\over 4}\bra u_\mu u_\nu\ket\bra u^\mu u^\nu\ket \right).
\end{eqnarray}

Comparing~(4.8) with the $P_i$ from~(2.4-5), the contributions of the
tensor-meson resonance to $\bar{l}_i$ are found to be
\begin{equation}
\begin{array}{rcl}
\bar{l}_1\,^T&=&\d -{32\pi^2 \over \gamma_1} \, {4K_T^2\over 3M^2} \;= \; -0.63
\\[4mm]
\bar{l}_2\,^T&=&\d {32\pi^2 \over \gamma_2} \, {4K_T^2\over M^2} \; = \; 0.95,
\end{array}
\end{equation}
with $\gamma_1={1\over 3}$ and $\gamma_2={2 \over 3}$. These results are
compatible with those obtained from another effective Lagrangian describing the
tensor-meson interaction with the pions developped in~\cite{Bell}.

\setcounter{equation}{0}
\section{Comparison of Pauli-Fierz scheme \newline with other ones}

This polynomial correction procedure to make a theory compatible with QCD can
be applied to the Pauli-Fierz type of Lagrangians, as may
be seen by using a description of a spin-1 field in their way and the
conditions developed in~\cite{Leut}. The Lagrangian
\begin{eqnarray*}
{\cal L}_V^{\rm kin}&=&\Bigl<{1\over 4}(V_{\mu\nu}V^{\mu\nu}-
2M_V^2V_\mu V^\mu)-\alpha\bar{C}^2+\beta^2\nabla_\mu\bar{C}\nabla^\mu\bar{C}
+\beta M_V\bar{C}V\Bigr> \\
\hspace{-1cm}\mbox{with }V_{\mu\nu}&=&{1\over \sqrt{2}}(\nabla_\mu V_\nu-
\nabla_\nu V_\mu),\\
V_S &=& \nabla_\mu V^\mu,\\
\bar{C} && \hspace{-1cm} {\rm  an \: auxiliary \: scalar \: field},\\
\alpha\neq 0 & {\rm and} & \beta\in\R,
\end{eqnarray*}
describes a spin-1 field of mass $M_V$ because
\[
\left(\begin{array}{c} V_S\\ \bar{C}\end{array}\right) =
  \left(\begin{array}{c} 0\\ 0\end{array}\right) \; \; \;
{\rm and} \; \; \; (\mbox{\large $\Box$} +M_V^2)V_{\mu}=0,
\]
can be derived from Euler-Lagrange equations in the same way as in Section~3.
The single condition $V_S=0$ does not fix all the parameters of
${\cal L}_V^{\rm kin}$, whereas for ${\cal L}_T$  the four conditions on
$\partial_{\mu}T^{\mu\nu}$ fix them all.

On the introduction of the couplings with the pseudoscalar mesons,
\[ {\cal L}_V^{\rm int} =\bra \bar{J}^{\mu\nu}V_{\mu\nu}+\bar{J}\bar{C}\ket,\]
the effective action becomes
\begin{eqnarray*}
S_V&=& {1 \over 2}\int{\rm d}x\,{\rm d}y\,\bra
\bar{J}^{\mu\nu}(x)\,\Delta_{\mu\nu\rho\sigma}(x-y)\bar{J}^{\rho\sigma}(y)+
\bar{J}(x) \bar{K}(x-y) \bar{J}(y)\ket\\ [4mm]
\hspace{-1cm}\mbox{with }\Delta_{\mu\nu\rho\sigma}(x)&=&\int{{\rm d}^4k\over
(2\pi)^4}{{\rm e}^{-{\rm i}kx}\over k^2-M_V^2+{\rm
i}\epsilon}\left[g_{\mu\rho}k_\nu k_\sigma-g_{\mu\sigma}k_\nu k_\rho-
(\mu\leftrightarrow \nu)\right]\\
\hspace{-1cm}\mbox{and } \bar{K}(x) &=& \delta^4(x) \left( {1\over 2\alpha}-
{3\beta^2\over 2\alpha^2} \mbox{\large $\Box$} \right).
\end{eqnarray*}
As in the tensor meson case, the coupling of $\bar{C}$ with the pions generates
polynomial terms that can be expressed in terms of the $P_i$ of (2.3).

This effective model {\it \`a la} Pauli-Fierz is different from the two
presented
in~\cite{Leut}. But imposing the same conditions to make it compatible with
QCD (the behaviour of the two-point functions, of the pion form factors and of
the forward amplitude for the elastic scattering of pseudoscalar mesons)
implies its equivalence with the two other descriptions (the antisymmetric
tensor and the vector field formulation) to $O(p^4)$.

\setcounter{equation}{0}
\section{Overview of the various contributions of the meson-resonances and
conclusion}

The roles played by different resonances can now be quantified. Following
from~(2.6) and~(4.9), we have
\begin{equation}
\bar{l}_i=-\ln\,{m^2\over \mu^2}+\sum_{R}\bar{l}_i\,^R,
\end{equation}
where \vspace{-2mm}

$\mbox{}$\qquad\begin{tabular}{l}
$m= 140$~MeV, is the pion mass,\\
$\mu=M_\rho=770$~MeV is the renormalization scale,\\
$\bar{l}_i\,^R$ is the contribution of the resonance $R$.
\end{tabular}

{}From~\cite{Eck}, the contributions to $\bar{l}_1$ and $\bar{l}_2$ of all the
meson-resonances of spin $\leq 1$ and of mass $\lsim$ $1$~GeV can be
extracted, with~(4.9) it gives:
\begin{eqnarray}&&\begin{array}{rcl}
\bullet\quad \mbox{vector:}\qquad \bar{l}_1\,^V&=&-4.5\\[2mm]
\bar{l}_2\,^V&=&2.2 \end{array}\\[4mm]
&&\begin{array}{rcl}\bullet\quad \mbox{scalar (singlet and octet):}\quad
\bar{l}_1\,^S&=&1.2\\[2mm]
\bar{l}_2\,^S&=&0\end{array}\\[4mm]
&&\begin{array}{rcl}\bullet\quad \mbox{tensor:}\quad
\bar{l}_1\,^T&=&-0.63\\[2mm]
\bar{l}_2\,^T&=&0.95.\end{array}
\end{eqnarray}

Thus, in absolute value, the vector-meson $\rho(770)$ contributions are about
twice as big as the sum of the others.

It must be noted that although $\rho'(1450)$ is a vector-meson with a mass
comparable to that of $f_2(1270)$, its contributions to the low energy
coupling-constants is only about $ {\Gamma^{\rho'}_{\pi\pi} \over
\Gamma^{\rho}_{\pi\pi}} {M_{\rho} \over M_{\rho'}}\cong {1 \over 20}$ of those
of $\rho(770)$, mainly because $\Gamma^{\rho'}_{\pi\pi}\cong {1 \over 10}
\Gamma^{\rho}_{\pi\pi}$~\cite{Par}.

We therefore assume that a good estimate of the contributions of the resonances
is obtained by considering their spectrum up to the tensor-meson mass. This
estimate can be compared with the experimental information:
\begin{eqnarray*}
\bar{l}_1\,^{\mbox{\tiny reson}}\cong -0.5 &\mbox{as}&
\bar{l}_1\,^{\mbox{\tiny exp}}=-0.7\pm 2.0\\[2mm]
\bar{l}_2\,^{\mbox{\tiny reson}}\cong 6.6 &\mbox{as}&
\bar{l}_2\,^{\mbox{\tiny exp}}=5.3\pm 1.3.
\end{eqnarray*}
The intervals for the phenomenological values given above cover all the
published values together with their
uncertainties~\cite{Bell,Pen,Rig,Beld,Bij},
at least to the best of our knowledge.

Besides this, the vector-meson cannot contribute to
$\bar{l}=2\bar{l}_1+4\bar{l}_2$ (low-energy combination involved in
$\pi^0\pi^0$ scattering, for instance), so that
\[\bar{l}\,^{\mbox{\tiny reson}}\cong 25.4 \quad\mbox{as}\quad
\bar{l}\,^{\mbox{\tiny exp}}=19.8\pm 9.2.\]

Thus it has been shown that although the tensor-meson resonance is 50\% heavier
than the dominating vector-meson one, it still contributes to the
renormalized coupling constants of chiral perturbation theory. This analysis
supports the statement that their phenomenological values may be understood on
the basis of the spectrum of low lying excited states.

\vspace{10mm}
It is a pleasure to thank H. Leutwyler for many useful discussions and a
critical reading of the manuscript and J. Gasser for valuable conversations.

\end{document}